\documentstyle[11pt,twoside,jltp,epsfig]{article}
\renewcommand{\v}[1]{{\bf #1}}

\newcommand{\gr}{{\nabla}}
\def\eqa{\begin{eqnarray}}
\def\eea{\end{eqnarray}}
\newcommand{\eq}{\begin{equation}}
\newcommand{\ee}{\end{equation}}
\newcommand{\nn}{\nonumber\\}
\newcommand{\Eq}[1]{Eq.~(\ref{#1})}

\newcommand{\ra}{\rightarrow}

\title{Inhomogeneity in doped Mott insulator}
\author{Dung-Hai Lee$^{1,2,3}$}

\address{${(1)}$Department of Physics,University of California,
Berkeley,  CA 94720, USA\\ ${(2)}$Material Science Division,
Lawrence Berkeley National Laboratory, Berkeley, CA 94720, USA
\\${(3)}$Center for Advanced Study, Tsinghua University, Beijing
100084, China} \runninghead{D.-H. Lee }{Inhomogeneity in doped
Mott insulator}

\begin{document}

\begin{abstract}
We introduce the concept that there are two generic classes of
Mott insulators in nature. They are distinguished by their
responses to weak doping. Doped charges form cluster (i.e.
distribute inhomogeneously) in type I Mott insulators while
distribute homogeneously in type II Mott insulators. We present
our opinion on the role inhomogeneity plays in the cuprates.
\end{abstract}

\maketitle

\section{INTRODUCTION}
Understanding of Mott insulators is one of the central problems of
condensed matter physics. In addition to electronic Mott
insulators recently an interesting Bose Mott insulator has been
produced by trapping bosonic neutral atoms in an optical
lattice\cite{bosemott}. The recent upsurge in interest in Mott
insulators is largely due to their dramatic response to doping.
For example by doping the transition metal oxides such as the
cuprates and the manganites, one obtains superconductors with the
highest known transition temperatures and the world's most
magnetoresistive materials.

Mott states, in addition to being insulating, can be characterized
by the presence or absence of a spontaneously broken symmetry
(e.g. spin antiferromagnetism), by the nature of the low energy
excitation spectrum (e.g. gapped or gapless)\cite{note}, and most
recently, by the presence or absence of topological order and
charge fractionalization\cite{bfg}. To this list, we add a
``type'' index that classifies Mott insulators into two types
depending on their response to doping.

In a type I Mott insulator an increasing chemical potential
induces a first order phase transition from an undoped state to a
charge-rich state so that the charge density changes
discontinuously. In other words, there is a range of ``forbidden
charge density,'' and two-phase coexistence. In a type II Mott
insulator, charges go in continuously above a critical chemical
potential. Depending on the ratio between the delocalization
energy and the interaction energy these doped charges can form a
Wigner crystal or a homogeneous liquid phase\cite{anothernote}. An
important implication of this work is the existence of a generic
class of Mott insulators which become inhomogeneous upon doping.

For the transition metal oxides discussed above, there exists
considerable evidence that light doping induces spatial
inhomogeniety\cite{ref1,stm,stm1,ref2,ref3}. Conversely, for Mott
insulator such as $Sr_{1-x}La_xTiO_3$\cite{lsto}, conventional
Fermi liquid behavior is observed for doping as low as $x=5\%$,
indicative of homogeneity.  Thus, the former are type I, the
latter is type II.

Generally speaking, insulating states with charge gap including
both band\cite{band} and Mott insulators occur in crystalline
systems at isolated rational ``occupation numbers,'' $\nu=\nu^*$,
where $\nu$ is the number of particles per chemical unit cell. By
``doping'' we mean a process which causes the occupation number to
shift away from $\nu^*$. When lattice translation symmetry is not
spontaneously broken, $\nu^*$ is typically an integer for bosons,
and an even integer for fermions with spin. The fermionic state
thus may be adiabatically connected to a weakly interacting
band-insulator, however a Bose Mott insulator is always a strong
correlation effect. Charge gapped insulating states can also occur
when $\nu^*$ is a rational fraction (for fermions this includes
odd-integer). Usually when that happens, translational symmetry is
spontaneously broken so that the unit cell of the reduced
translation group has integral $\nu$ for
bosons\cite{leesh,oshikawa} and even integral $\nu$ for fermions.
For instance, electronic Mott insulators with $\nu^*=1$ often
exhibit antiferromagnetic N{`e}el long-range order, which doubles
the unit cell leading to an effective $\nu_{eff}=2$. Nevertheless
there exists model bosonic systems for which the Mott  state can
be shown to have no broken symmetries for
$\nu^*=1/2$\cite{rk,sondhi,bfg} . (Currently,  no laboratory
system has been found which unambiguously exhibits this exotic
behavior.)

The rest of this paper addresses the response of Mott insulators
to light doping (i.e. $\nu\ra\nu^*-\epsilon$). Our central
observation is that regardless of their classification (i.e.
symmetry, gap, topology) Mott insulators can be divided into two
groups differentiated by whether they remain homogeneous after
doping\cite{coulomb}. Furthermore we shall show that these two
types of insulating state is analogous to the two types of
superconducting state under magnetic field. In particular if the
Mott insulators are two dimensional and the constituent particles
are bosons there exists a mathematical mapping, the so-called
``duality transformation'', that relates their zero-temperature
response to doping to the finite-temperature response of a 3D
superconductor to a magnetic field\cite{leesh,fl}. (The duality
transformation has been used to deduce two types of doping
behavior by Balents {\it et al} in the context of doping a spin
liquid called ``nodal liquid''\cite{bfn})The following table
summarize correspondence between the two.

\begin{tabular}{|l|l|} \hline
{\it $T=0$ properties of 2D Bose Mott insulators} & {\it $T>0$ properties of 3D Superconductors} \\
\hline\hline Doping & Applying magnetic field \\ \hline Chemical
potential $\mu$ & Applied magnetic field H\\ \hline Induced particle density $\rho$ & Magnetic induction $B$ \\
\hline World line of doped particles & Flux tubes\\ \hline Quantum
delocalization of doped particles & Thermal
meandering of flux tubes\\ \hline Type I Mott insulator & Type I superconductor \\
 Mott gap & $H_c$\\ Effective attraction between doped particles &
Positive N-S interface energy\\  \hline Type II Mott insulator &
Type II superconductor\\ Effective repulsion between doped
particles & Negative
N-S interface energy\\
 Mott gap & $H_{c1}$ \\ Wigner crystal of
doped particles &Abrikosov flux lattice \\
Superfluid state & Entangled vortex fluid\\
Critical $\mu$ at which Wigner crystal melts &$H_{c2}$\\
\hline
\end{tabular}
\section{DOPING A TRANSLATIONALLY INVARIANT BOSE MOTT INSULATOR}

In order to avoid the issue of spontaneous symmetry breaking in
electronic Mott insulators and focus on the absolute essentials,
we consider the simplest kind of Mott insulators - the ones formed
by spin zero point bosons on a lattice. Due to Ref.\cite{bosemott}
this consideration is no longer an academic exercise\cite{dope}.
Consider the following Hamiltonian \eqa H&&=-{t\over 2}\sum_{<ij>}
( a_i^+a_j+ h.c.) + {U\over 2} \sum_i (a_i^+a_i)(a_i^+a_i-1)\nn&&+
{1\over 2}\sum_{i,j} V_{ij} a_i^+a_ia_j^+a_j\
-\mu\sum_ia^+_ia_i,\label{model} \eea where $i,j$ label the
lattice sites on a D-dimensional hypercubic lattice, and $a_i^+$
creates a boson at site $i$. The first term of \Eq{model}
describes the quantum mechanical ``hopping'' of bosons from a site
$i$ to its nearest
neighbors $j$, 
the second and the third terms  describe the pair-wise interaction
between bosons. The $U$ term is a contact interaction and the
$V_{ij}$ terms describe interaction between bosons separated by
$|\v r_i-\v r_j|$.

Let us focus on the limit where $U/t \to \infty$.
In this ``hardcore'' limit  doubly occupied site costs  energy
$U$, hence are excluded.
In this case for large positive $\mu$ there is an unique ground
state
\eq |{\rm Mott}>=\prod_j a_j^+|0>, \label{state} \ee in which each
site is occupied by one and only one boson and hence $\nu=1$.
Since this state is separated from all other states with the same
particle number by an energy gap of order $U$, clearly we have an
insulator.

The behavior of the system upon decreasing $\mu$ (i.e. doping)
depends on the values of $V_{ij}$. For $V_{ij}=0$ the doped holes
(i.e the empty sites) interact only through the hardcore
exclusion. For appropriate $\mu$ where the ground state has a
small $1-\nu$, the hopping term is clearly minimized by
delocalizing the holes around, since there is plenty of room
available for them. In fact it is known that such system is an
uniform superfluid with superfluid stiffness proportional to
$t(1-\nu)lnln[1/(1-\nu)]$\cite{dsf}. Conversely, if $V_{ij}$ is
attractive (negative) with range $|\v r_i-\v r_j|\le R$ it is
clear that the holes will cluster when the strength of this
interaction is strong compared with $t$ - the holes will phase
separate.
Thus depending the value of $V_{ij}$ \Eq{model} can describe
either a type I or type II Mott insulator when $\nu=1$.

Consider, for example, the case in which $V_{ij}=-V$ for $(ij)$
nearest-neighbor sites, and $V_{ij}=0$ otherwise.  Technically in
this limit \Eq{model} is equivalent to the S=1/2 ferromagnetic XXZ
model in a z-direction magnetic field \eq H =
-J_{xy}\sum_{<ij>}(S_i^xS_j^x+S_i^yS_j^y) - J_z
\sum_{<ij>}S_i^zS_j^z - h_z\sum_i S_i^z.\label{xxz}\ee The mapping
between these two models relates $J_{xy}$ to $t$, $J_z$ to $V$,
$h_z$ to $\mu+Vc/2$ ($c=$ the coordination number). The $z$
component of the magnetization is related to the boson density
according to $M_z=(\nu-1/2)N$. (Here $N$ is the total number of
lattice sites.) For $J_z<J_{xy}$ (i.e.$V<t$), this model has XY
order in the absence of $h_z$. In this range of parameters varying
$h_z$ causes the magnetization $M_z$ to vary continuously. Thus
for each fixed $M_z$ the ground state is uniform and
ferromagnetic. In terms of bosons this means that the doped system
is a uniform superfluid for all $\nu$. Conversely, for
$J_z>J_{xy}$ (i.e. $V>t$) the model is effectively an Ising
ferromagnet with fully polarized ground state. In this case $M_z$
exhibits a discontinuity $\Delta M_z=N$ at $h_z=0$. Thus for all
$|M_z|\ne 1$ the ground state exhibits phase separation into two
oppositely polarized domains. In terms of the bosons, these two
domains are Mott insulating ($\nu=1$) and empty ($\nu=0$)
respectively.

\section{DUALITY TRANSFORMATION FOR THE BOSE MOTT INSUALTOR IN D=2}

 While the above discussions applies to arbitrary space
dimensions, there is a particularly convenient way of thinking
about Bose Mott insulators that  is specific to $D=2$. It turns
out that for the class of model given by \Eq{model},
there is a mathematical mapping, the ``duality'' transformation,
that provides us an alternative view of the physics of \Eq{model}
in terms of the vortices of the boson field. It is this mapping
that enables us to establish a precise connection between the two
types of Mott insulators with type I and type II superconductors
\cite{fl,leesh,bfn}. In the following we discuss the physical
content of the duality transformation without going into its
technical details of the tranformation\cite{leesh}.

A vortex is a topological defect in the Bose field. When a boson
is adiabatically transported around a vortex, the boson
wavefunction acquires a phase factor - the Aharonov-Bohm phase. In
the dual picture, the vortices are viewed as particles (they turn
out to have  Bose statistics as well), and when they are brought
around an original boson they acquire an Aharonov-Bohm phase.  The
fact that  a boson and a vortex  acquire a phase when they go
around one another implies that bosons and vortices can not Bose
condense simultaneously. As is well known, in the Bose superfluid
phase the vortex density fluctuation must be absent. Conversely,
in the dual phase, where the vortices form a superfluid, the boson
density (and hence the dual magnetic flux) must be frozen;  the
vortex superfluid phase is the Mott insulating phase of the
original bosons\cite{note10}.
It is important to note that the absence of boson density
fluctuation is a necessary but not sufficient condition for vortex
condensation. For example a {\it static} boson density can still
frustrate vortex condensation because it acts like a background
magnetic field. However when the static boson density corresponds
to an integral $\nu$, the vortices see a background magnetic flux
corresponding to integral number of flux quanta per plaquette.
(The vortices live on the dual lattice, i.e., the centers of the
square plaquettes.) This type of flux is ``invisible'' to the
vortices because they can be ``gauged away''. The $\nu=1$ Bose
Mott insulator, discussed in the previous section, corresponds to
precisely this situation.

When the boson density is a fraction ($\nu=p/q$) it is also
possible for the vortices to condense. Such a state is most
naturally accompanied by spontaneous translation symmetry
breaking, as discussed above, leading to an enlarged unit cell
with an effective integer $\nu$\cite{leesh}. However, it is
possible to imagine a more exotic Mott state at $\nu=p/q$ in which
the translation symmetry is unbroken. This could happen if $q$
elementary vortices form a bound-state, and these composites then
condense. Such a composite condensation is unfrustrated by an
unform static boson density $\nu=p/q$\cite{bfg}. Moreover, since
the condensate vortex consists of $q$ elementary vortices, the
charge $1/q$ bosonic soliton excitation (viewed by the condensate
vortices as a flux quantum) can become a finite energy excitation.
Thus there is fractional charge solitons!

In short, a vortex condensate requires the bosons to Mott
insulate.
Consequently we can view a boson Mott insulator as a vortex
superconductor. 
Doping changes the average background boson density. To the
vortices this appears as a change in the background magnetic
field. In this way the doping properties of the boson Mott
insulator is related to the magnetic properties of the vortex
superconductor.


\section{THE ANALOGY WITH SUPERCONDUCTORS IN A MAGNETIC FIELD}

In the above discussion a zero temperature boson Mott insulator is
mapped onto a zero-temperature vortex superconductor (with quantum
fluctuating two-dimensional electromagnetic fields)\cite{note3}.
The final step is to realize that the quantum partition function
of a (particle-hole symmetric) two dimensional superconductor with
fluctuating gauge field is equivalent to the classical (i.e.
thermal) partition function of a three-dimensional superconductor
with thermally fluctuating magnetic field\cite{note4}.  This final
correspondence between the Mott insulator and the classical
fluctuating 3D superconductor is summarized in table I.

A lot is known about thermally fluctuating 3D superconductors.
Mean-field theory predicted that there are two types. For a type I
superconductor the magnetic induction $B$ jumps discontinuously
from $B=0$ to $B=H$ at $H=H_c$. As the result there is a range of
$B$ (i.e. $0<B<H_c$) in which phase separation occurs. For example
one way of enforcing a fixed average magnetic induction is to
place a flat slab of type I superconductor under magnetic field
$H<H_c$. It is known that when that is done ``intermediate state''
where superconducting region and normal region alternate occurs.
It is interesting that among many possible inhomogeneous
structures the laminar structure (or stripe) has been
observed\cite{tink}. Translate this using table I we conclude that
a type I Bose Mott insulator undergoes a first order insulator
$\ra$ superfluid transition as a function of chemical potential.
The density of doped bosons jumps discontinuously at the
transition. When the doping density is fixed at a value smaller
than the critical density at the first order transition the system
phase separates.

Mean-field theory predicts that in a type II superconductor the
magnetic induction $B$ increases from zero continuously at the
lower critical field $H_{c1}$. For $H>H_{c1}$ the magnetic
induction appear in the form of flux tubes each enclosing a single
quantum of magnetic flux. These flux tubes form a regular lattice
in the absence of disorder. Unlike the type I superconductors
thermal meandering of the flux tubes can affect the physics of
type II superconductors dramatically near $H_{c1}$ and $H_{c2}$.
For example near $H_{c1}$ the distance between neighboring flux
tubes is much greater than the range of their interaction (the
London penetration depth $\lambda$). As the result thermal
meandering of the flux tubes can melt the flux lattice into flux
liquid\cite{melt}. At larger magnetic field the density of flux
tube becomes higher so that their interaction can stabilize the
flux lattice. This flux lattice persists until $H\ra H_{c2}$ where
the thermal meandering melt the flux lattice again. Translating
the above using table I implies that the extra carriers enter a
type II Mott insulator continuously at a critical chemical
potential. Once they enter they can either delocalize (hence Bose
condense) or form a Wigner crystal depending on the relative
importance of the delocalization energy and the interaction
energy. At at very low carrier density the delocalization always
win and we expect the boson hopping to render the system a
superfluid.

What determines a superconductor to be type I or type II is the
ratio between the London penetration depth and the core size of
the vortices, or more physically in terms of the sign of the
interface energy between the normal and superconducting regions.
Type I superconductors have a positive interface energy while type
II superconductors have a negative one. As the result the flux
tubes effectively attract each other in type I superconductors
while repel each other in type II superconductors. As we have seen
this is exactly how we turn a type I Mott insulator into a type II
one in \Eq{model}.

\section{DOPING A BOSE MOTT INSULATOR WITH AN ORDER PARAMETER}

As we discussed at the beginning of this paper a Mott insulating
state can be accompanied by translation symmetry breaking. For
example let us consider \Eq{model} with $U/t\rightarrow \infty$
and $V_{ij}=+V$ for nearest neighbor $<ij>$ and $0$ otherwise. For
sufficiently strong $V$ and $\mu=0$ the ground state breaks
translation symmetry and bosons form a checkerboard lattice and
Mott insulate. In this two-fold degenerate ground state the unit
cell is doubled. Is this Mott insulator type I or type II?

With the above specific choice of $U$ and $V_{ij}$ \Eq{model} is
equivalent to \Eq{xxz} with $J_{xy}=-t$, $J_z=V$ and $h=\mu-Vc/2$.
For this choice of parameters it is known that as a function of
$h_z$ \Eq{xxz} exhibits a ``spin flop'' transition from the
antiferromagnetic Ising ($S_z$) ordered phase into the
ferromagnetic XY ordered phase. Translate this into the boson
language it implies that the boson Mott insulator is type I.

Before closing this section we argue that the existence of an
order parameter ($\psi$) in the insulating state (for the above
example $\psi$ is the two-sublattice density wave order parameter)
has an effect in determining the type of the Mott insulator. If we
assume that doping changes the value of $\psi$, then at the
interface between doped and undoped region a spatially varying
$\psi$ is necessary. Through the spatial gradient energy \eq \int
d^dx |\gr\psi|^2\ee  a positive contribution to the surface energy
is resulted. Consequently the presence of an order parameter in
the insulating state drives the system toward a type I Mott
insulator.

\section{What ROLE DOES INHOMOGENEITY PLAY IN THE CUPRATES;
PAIRING IN THE HOLE-RICH ISLANDS IN $Bi_2Sr_2CaCu_2O_{8+x}$}

Finally we come to address the question {\it what role does
inhomogeneity play in the curpates?} One of the most direct
evidence of electronic inhomogeneity in the cuprates comes from
the STM image of the surface of $Bi_2Sr_2CaCu_2O_{8+x}$, where
nanoscale spatial variation of the energy gap in the tunnelling
density of states is observed\cite{stm,stm1}. (Of course we have
to assume that the STM results is not a surface artifact.) As
emphasized by Lang {\it et al}\cite{stm1}, the characteristics of
the tunnelling spectra divide into two distinct types: ``$\alpha$
region'' where sharp coherence peaks exist ($\Delta$ ranges from
25 -50 meV), and ``$\beta$'' region where there is no sharp
coherence peak ( $\Delta$ ranges from 50 to ~75 meV). We take
these evidences as suggesting that in the $\alpha$ region
superconducting pairing has clearly won over all other competing
orders, while in the $\beta$ region it has not. We believe that
inhomogeneity allows the superconducting $\alpha$ region to exist
in the underdoped regime. As the result superconductivity prevails
over a wider range of doping concentration. However, we do not
think inhomogeneity plays an important role in the pairing of
$\alpha$ regions. The above line of thinking suggests that in
order to understand the pairing mechanism it is best to
concentrate on the $\alpha$ regions.

Since the superconducting gap can vary over the length scale of a
few nanometers, the coherence length must not be significantly
longer than that. At such length scale (comparable with the
averaged inter-hole distance) the Coulomb interaction is poorly
screened. How can pairing tolerate such strong unscreened Coulomb
interaction?

In the rest of this section we present an explicit example of a
paired state that can survive strong repulsion. The state we shall
concentrate on is of the form \eq
|\psi(\Delta)>=P({n_i})|\psi_{dBCS}(\Delta)>.\label{jestrow}\ee In
\Eq{jestrow} $|\psi_{dBCS}(\Delta)>$ is a d-wave BCS state
characterized by the gap parameter $\Delta$. The operator $P$ is a
Jestrow correlator that suppresses configurations with large
density fluctuations. A particular form of $P$, that is often used
in the high $T_c$ context, is the Gutzwiller projection operator.
In our opinion although we do not know precisely what Hamiltonian
gives rise to a ground state like \Eq{jestrow}, it is rather safe
to assume that the ground state in the $\alpha$ region is of that
form.

In the following we demonstrate that a state like \Eq{jestrow} has
the property that it is more sensitive to magnetic interactions
than to charge interactions. To do that we consider the following
Hamiltonian \eqa H&=&-t\sum_{\langle ij\rangle} (c_{j\alpha}^\dag
c_{i\alpha} \!+h.c.) +J\sum_{\langle ij \rangle} (\v S_i\cdot\v
S_j \! -\frac{1}{4}\!n_{i}n_{j})\nonumber \\
 &+& {V_c}\sum_{i>j}
 {1\over {r_{ij}}}(n_{i}-\bar{n})(n_{j}-\bar{n}).
\label{h} \eea The purpose of the following discussions is not to
prove \Eq{jestrow} is the ground state of \Eq{h}. Rather we hope
to demonstrate one point - the state given in \Eq{jestrow} is more
sensitive to the magnetic interaction $J$ than to the charge
interaction $V_c$. The result we shall quote below is obtained by
Wang and collaborators in Ref.[26].


\begin{figure}
\epsfxsize=8cm \epsfysize=8cm \vskip -1.8cm \epsfbox{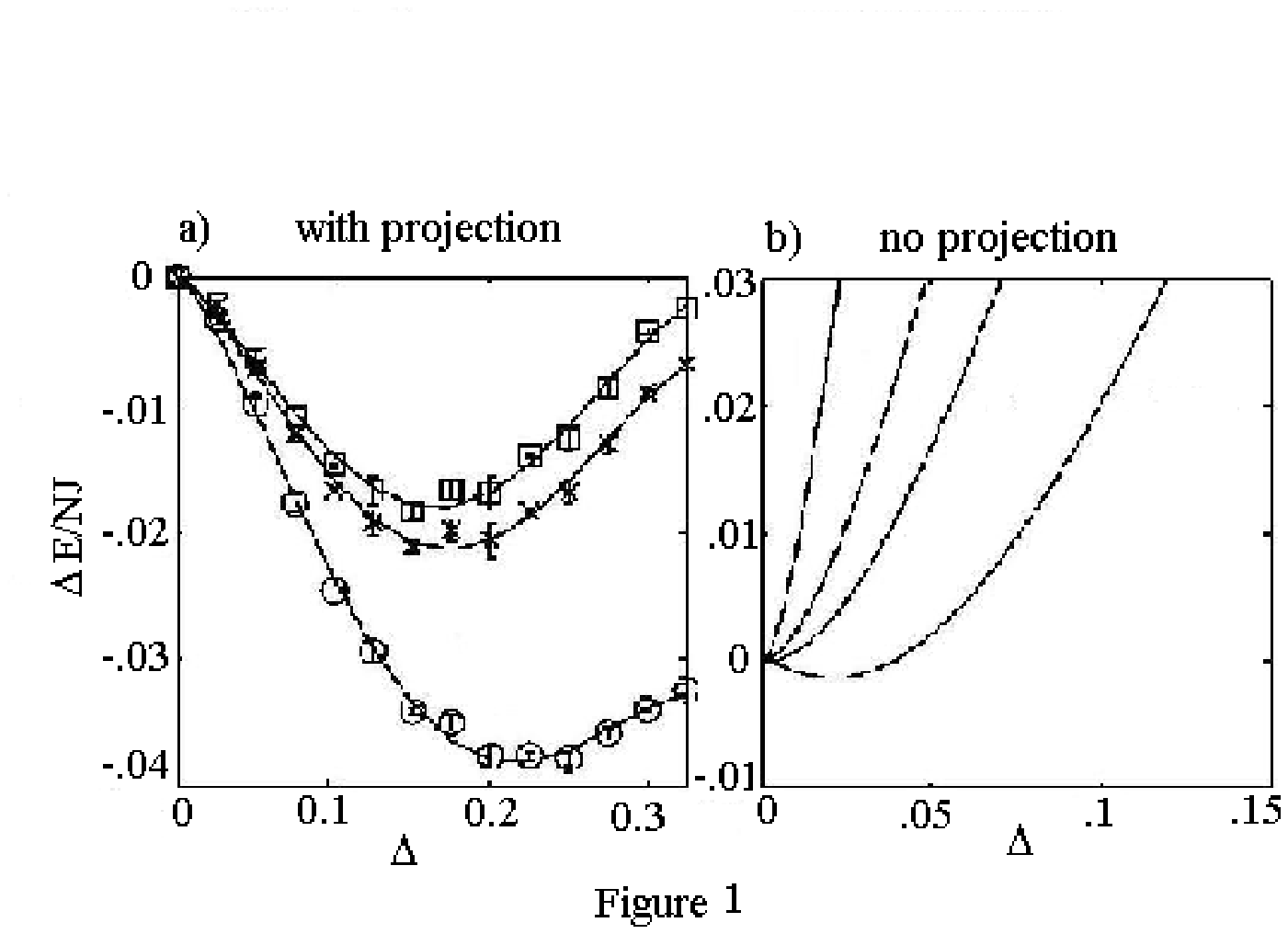}
\caption{\small Results for $x=12$\%. $\Delta E$ as a function of
$\Delta$ from Gutzwiller projection in a $10\times 10$ lattice.
Open circles: pure t-J model; crosses: t-J model with nearest
neighbor repulsion $V_{nn}=3J$; squares: t-J model with Coulomb
interaction $V_c=3J$.}
\end{figure}

By a straightforward Monte-Carlo minimization Wang {\it et al}
conclude that for doping $x=0.12$ it is energetically favorable to
develop a non-zero $\Delta$ for $V_c$ as big as $9J$. To
appreciate the effects of the Jestrow correlator $P$ (which they
take as $P=\prod_i (1-n_{i\uparrow}n_{i\downarrow}$) in
\Eq{jestrow} they compared the results with and without $P$.
Interestingly without $P$ a nearest-neighbor repulsion $V_{nn}$
destabilizes pairing when $V_{nn}$ is larger than $\approx 0.5J$!
 Thus the Jestrow correlator $P$ enhances the
stability of pairing in \Eq{jestrow} by as much as a factor of 18!

The following is a brief summary of Wang {\it et al}'s results.
Given \Eq{jestrow} Wang {\it et al} minimize
$E(\Delta)=\langle\Psi|H|\Psi\rangle/\langle\Psi|\Psi\rangle$ by
varying $\Delta$. The results presented below are obtained for
$x=0.12$ in $10\times 10$ lattice by  variational Monte-Carlo.

In Fig.1(a) $\Delta E\equiv E(\Delta)-E(0)$ versus $\Delta$ is
plotted. The wavefunction used is \Eq{jestrow} with $P$ being the
Gutzwiller projector. The open circles are for the pure t-J model
(i.e. without charge-charge interaction), the crosses are for t-J
model with a nearest neighbor repulsion $V_{nn}=3J$, and the open
squares are for t-J + Coulomb model (\Eq{h}) with $V_c=3J$. For
each of the three cases a nonzero $\Delta$ develops.

In Fig.1(b) $\Delta E\equiv E(\Delta)-E(0)$ versus $\Delta$ is
plotted for the repulsive nearest-neighbor ($V_{nn}$) model. An
important difference is that the Jestrow correlator $P$ in
\Eq{jestrow} is {\it removed}. This time the optimal $\Delta$
vanishes for $V_{nn}\geq 0.5J$.

These results clearly indicate that a strongly correlated paired
state such as \Eq{jestrow} can indeed sustain strong repulsion
between the electrons. In addition we take this as suggesting a
state like \Eq{jestrow} is more sensitive to magnetic interactions
than the charge interactions.

In the literature it is often stated that pairing correspond to
real space binding of holes. In the presence of strong Coulomb
interaction real space bound hole pairs are extremely
energetically unfavorable, and indeed in an ansatz such as
\Eq{jestrow} no hole binding is present.

\section{CONCLUSIONS}

To summarize, in this paper we introduce the concept that there
are two types of Mott insulator in nature. A type I Mott insulator
becomes inhomogeneous after doping, while a type II Mott insulator
remains homogeneous. We present a specific lattice boson models
which Mott insulates without symmetry breaking at $\nu=1$ and,
depending on the sign of a microscopic interaction, exhibits these
two types of ground states after doping. We argue that the
presence of a symmetry breaking order parameter in the insulating
state has the effect of driving the system toward type I. In
addition, we argue that these two types of Mott insulating states
are dual to the type I and type II superconducting states. Here we
conjecture that the notorious $x=0.19$ ``quantum critical point''
in the cuprates is related to the inhomogeneous to homogeneous
transition when the magnetic induction ($B$) in a type I
superconductor is varied across $H_c$.

We also express our opinion on the role played by the
inhomogeneity in the cuprates. To reiterate, we do not think
inhomogeneity plays a central role in pairing. However it does
allow the superconducting region (where pairing dominates over
other competing orders) to protrude into the underdoped regime. As
the result superconductivity prevails over a wider range of doping
concentration. Finally we believe a correlated pairing state given
by \Eq{jestrow} can describe the hole-rich superconducting regions
(the so-called $\alpha$ regions) imaged by the STM. Using the
result of Ref.[26] we demonstrate that this state is much more
sensitive to spin-spin interaction than charge-charge interaction.
This provides an mechanism by which pairing can survive poorly
screened Coulomb interaction.

\section*{ACKNOWLEDGMENTS}
Part of this paper (two types of Mott insulator) is the result of
a collaboration with S.A. Kivelson. I thank Seamus Davis and
members of his group for valuable discussions. DHL is supported by
NSF grant DMR 99-71503.



\begin{thebibliography}{99}
\bibitem{note} A Mott insulator necessarily has a charge gap, however it is
possible for it to have a gapless spin excitation spectrum.
\bibitem{band} A band insulator insulates because of
Pauli exclusion principle. In the non-interacting prototype of
this type of insulator all available single-particle orbitals are
occupied. This type of insulator usually possesses a even number
of electrons per crystalline unit cell.
\bibitem{coulomb} In the presence of long-range Coulomb forces macroscopic charge
inhomogeneities is impossible as $\nu-\nu^*\rightarrow 0$.
However, so long as the long-range forces are not too strong,
there could exist a range of low doping in which the doped charges
tend to cluster. We still refer to this kind of system as type I
Mott insulators as well. 
For instance, if a low density of doped charges form puddles or
stripes, we consider this a form of micro-phase-separation,
while a Wigner crystal of doped charges is deemed homogeneous.
This is because in a Wigner crystal charges localize in the
smallest unit hence no clustering is exhibited.
\bibitem{dope} The system considered in this Ref.\cite{bosemott} can  be ``doped'' by varying the period of the trap lattice.
\bibitem{note10} We note that the vortex superfluid discussed
here differs from the generic boson superfluid because the net
vorticity is zero. The field theory that describes such
"particle-hole symmetric" condensate has space and time on equal
footing (i.e. relativistic). This fact makes the later analogy
with thermal fluctuating 3D superconductors completely
appropriate.
\bibitem{note3} The field strength $F_{\mu\nu}$ ($\mu=0,1,2$) of a 2+1-dimensional EM field has three independent components.
The correspondence of the these three components with the boson
three-current $(\rho,\v j)$ is as follows: $F_{12}\leftrightarrow
\rho, F_{01}\leftrightarrow j_2, F_{02}\leftrightarrow j_1$.
\bibitem{note4} In the correspodence between a quantum superconductor at $T=0$ and a thermally fluctutating
superconductor at finite temperature the magnetic field ($F_{12}$)
and electric fields ($F_{01},F_{02}$) in the 2+1 dimensional
quantum theory is mapped on to $B_z$ and $B_y, -B_x$ respectively.
\end{thebibliography}
\end{document}